\documentclass[sigconf]{acmart}
\usepackage{wrapfig,lipsum,booktabs}
\usepackage{sidecap}
\usepackage{wrapfig}
\usepackage{pgf}
\usepackage{tikz}
\usepackage{amsmath}
\usepackage{amssymb}
\usepackage{times}
\usepackage{array}
\usepackage{xspace}
\usepackage{verbatim}
\newcolumntype{L}[1]{>{\raggedright\let\newline\\\arraybackslash\hspace{0pt}}m{#1}}
\newcolumntype{C}[1]{>{\centering\let\newline\\\arraybackslash\hspace{0pt}}m{#1}}
\newcolumntype{R}[1]{>{\raggedleft\let\newline\\\arraybackslash\hspace{0pt}}m{#1}}

\usepackage{balance}
\usetikzlibrary{arrows,automata}
\usepackage{hyperref}
\usepackage{rotating}
\usepackage{amsmath}
\usepackage{multirow} 
\usepackage{subfigure}
\usepackage{graphicx}
\usepackage{tabularx}
\usepackage{booktabs}
\usepackage{amssymb}
\usepackage{enumerate}
\usepackage{xcolor}
\usepackage{color}
\newcommand{\BfPara}[1]{{\noindent\bf#1.}}

\renewcommand{\a}{{\bf a}}

\newcommand{\x}{{\bf x}}

\usepackage{xspace}
\usepackage{setspace}
\newcommand{\ngram}{$n$-gram\xspace}
\newcommand{\chatter}{{Chatter}\xspace}
\newcommand{\automal}{{AutoMal}\xspace}
\newcommand{\knn}{{$k$-NN}\xspace}

\begin{document}

%\runningheads{A. Mohaisen, O. Alrawi, J. Park, J. Kim, D. Nyang, M. Mohaisen}{Network-based Analysis and Classification of Malware using Behavioral Artifacts Ordering}

\title{Network-based Analysis and Classification of Malware using Behavioral Artifacts Ordering}

\author{Aziz Mohaisen}
\affiliation{%
  \institution{University of Central Florida}
}
%\email{mohaisen@ucf.edu}
\author{Omar Alrawi}
\affiliation{%
  \institution{Georgia Institute of Technology}
}%\email{}
%\email{marxy@uho.edu}
\author{Jeman Park}
\affiliation{%
  \institution{University of Central Florida}
}%\email{}

\author{Joongheon Kim}
\affiliation{%
  \institution{Chung-Ang University}
}%\email{}
\author{DaeHun Nyang}
\affiliation{%
  \institution{Inha University}
}%\email{}
\author{Manar Mohaisen}
\affiliation{%
  \institution{University of Central Florida}
}%\email{manar.subhi@gmail.com}

%\author{Aziz Mohaisen, Omar Alrawi, Jeman Park, Joongheon Kim, DaeHun Nyang, Manar Mohaisen}

%\address{\affilnum{1}University of Central Florida\\
%\affilnum{2}Georgia Institute of Technology\\
%\affilnum{3}Chung-Ang University\\
%\affilnum{4}Inha University\\
%\affilnum{5}Korea University of Technology and Education}

%\institute{
%	Aziz Mohaisen \at
%	mohaisen@ucf.edu
%	\and	
%	Omar Alrawi \at
%	alrawi@gatech.edu 
%	\and 
%	Jeman Park \at
%	parkjeman@knights.ucf.edu 
%	\and 
%	Joongheon Kim \at
%	joongheon@cau.ac.kr 
%	\and 
%	DaeHun Nyang \at
%	nyang@inha.ac.kr
%	\and
%	Manar Mohaisen \at
%	manar.subhi@gmail.com
%}

\begin{abstract}
Using runtime execution artifacts to identify malware and its associated ``family'' is an established technique in the security domain. Many papers in the literature rely on explicit features derived from network, file system, or registry interaction. While effective, the use of these fine-granularity data points makes these techniques computationally expensive. Moreover, the signatures and heuristics are often circumvented by subsequent malware authors. In this work, we propose \chatter, a system that is concerned only with the \textit{order} in which high-level system events take place. Individual events are mapped onto an alphabet and execution traces are captured via terse concatenations of those letters. Then, leveraging an analyst labeled corpus of malware, $n$-gram document classification techniques are applied to produce a classifier predicting malware family. This paper describes that technique and its proof-of-concept evaluation. In its prototype form only network events are considered and eleven malware families are used. We show the technique achieves 83\%-94\% accuracy in isolation and makes non-trivial performance improvements when integrated with a baseline classifier of combined order features to reach an accuracy of up to 98.8\%.
\end{abstract}


%\fnotetext[1]{Corresponding author.  Email: \email{mohaisen@ucf.edu}}

\maketitle

\section{Introduction}\label{sec:introduction}

Malware has emerged as a challenging threat with the increased infection rates and levels of sophistication~\cite{govermentattacks,nissan,ramilli2010multi}. Examples of such threats include data exfiltration~\cite{wu2012whispers}, denial-of-service attacks~\cite{WangMCC15b}, and espionage~\cite{li2011evidence}, among many others. In order to defend against malware threats, the research community has spent a tremendous amount of efforts understanding their behavior so that detection, classification, and labeling of malware are performed with high accuracy using analysis techniques~\cite{mohaisen2013towards,MohaisenA14b,CaseldenBPMS13,YinSEKK07,maldimva,malkdd,rieck2008learning}. Threat information sharing for the improvement of the malware detection system by collecting and sharing more information also has been actively studied~\cite{Kampanakis14,Park18AAKKNM,Tosh15SKKM}.

Malware analysis aims to inspect binaries in various ways, and techniques fall mainly into two categories: static and dynamic analysis.  Static analysis is achieved without running the malware and consists of examining meta-data and patterns in the binaries. On the other hand, dynamic analysis utilizes artifacts that are generated by malware at runtime. The process for dynamic analysis entails executing a malware sample in a controlled environment so that changes in the environment including network traces, file system changes, memory access, and registry modifications, are all observed and recorded. The advantages of static analysis are multifold: static analysis techniques allow for comprehensive code examination of malware and are considered fast and scalable because they look for pre-computed signatures in the binaries. However, attackers can evade static analysis techniques through code obfuscation. In contrast, dynamic analysis is capable of detecting unseen malicious behavior contained in obfuscated code, but it requires setting up the sandbox environment and allocating resources for execution, which is an expensive process. 

Typically, execution of a fully-fledged analysis system such as AMAL~\cite{MohaisenA15} requires observing interactions between a binary and the underlying system's components, such as memory, file system, registry and network, for a certain amount of time to extract behavioral artifacts and build a feature vector from which the malware family is then identified. This, in turn, requires analyzing 100s of megabytes of memory, various gigabytes of virtual file system space, several megabytes of the registry, and large network traces of packet capture. 

To reduce the overhead of the dynamic analysis, while benefitting from its advantages over static analysis, we may utilize either of two approaches. First, instead of extracting deep features from the various classes of artifacts, we may only use shallow features. For example, instead of considering types of DNS responses, types of DNS records, counts per type, features of queried name servers, and so forth, we may only consider the total number of queried domains and returned responses, among others as rather a shallow feature. Similarly, instead of looking for deep content-based memory features, we may consider the shallow feature of counts over memory access. We notice that, while this approach is promising in reducing the overhead of dynamic execution it is only limited: shallow features are easy to circumvent to evade detection, and much of the overhead is consumed in executing the malware, rather than analyzing individual artifacts (e.g., online feature extraction can be done using API hooking~\cite{DinaburgRSL08}).

\begin{figure}[t]
\begin{center}
\includegraphics[width=0.4\textwidth]{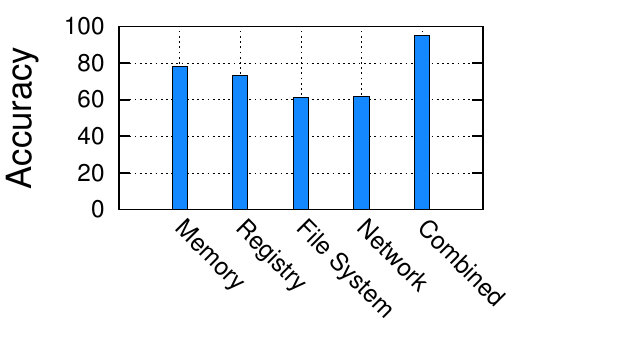}
\caption{The accuracy of classifying malware (ShadyRAT) based on various classes of artifacts in isolation and combined (using AMAL~\cite{MohaisenA15}, a fully-fledged behavior-based malware analysis system).}\label{fig:motivation}
\end{center}
\end{figure}

Another approach to reduce the overhead is by focusing on a certain class of artifacts, instead of multiple classes. We notice that not all classes of features are equally important contributors to the accuracy of malware classifiers. Consider the classification results shown in \autoref{fig:motivation} of the malware family Shady RAT (simply ShadyRat, or SRAT), and the contribution of various behavioral artifact classes in isolation and combined; memory, registry, file system, and network. We notice that, while the combined features extracted from the four different classes provide an accuracy of more than 95\%, the accuracy per individual class of features varies, and ranges from the mid-50\% to 75\%, indicating that not all classes of features are equally important. This also suggests that a single class of behavior has the potential of being used independently for classifying malware. 

However, we note two issues with even with the best class of features (network). First, it does not provide high enough accuracy to warrant its use independently. Second, it is unclear how difficult (and perhaps easy) it would be for an adversary to circumvent a set of features that rely on a single class of behavioral artifacts. To this end, the objective of this work is to look into a new type of features engineered from a single class of behavioral artifacts to boost the accuracy of classification and to be robust against arbitrary evasions by the adversary.

Determining the correct abstraction level at which to derive features is equally important and yet very challenging~\cite{malkdd,maldimva,MaricontiOACRS16,aafer2013droidapiminer,jang2011bitshred}. Features that accurately represent malware families are an important and defining criterion for high-fidelity malware classification. Furthermore, the level of complexity needed to obtain such abstractions in operational settings determines the potential of adopting and accepting such systems at scale. For example, sandboxing and virtual execution is a privilege that comes with high costs in online detection systems~\cite{RossowDGKPPBS12}. 
When a piece of malware runs on a host, collecting deep features becomes invasive to users on those hosts. To this end, while obtaining indicative artifacts and features is very important, the degree of invasiveness is also a significant consideration for operational systems. 

In this paper, we introduce \chatter, a  behavior-based system for collecting run-time artifacts and feature analysis and derivation, to address some of those problems. \chatter is a robust system that relies on the order of the behavioral artifacts that malware samples generate. \chatter can be implemented as a stand-alone system or integrated into more robust execution-based systems. As its ground truth, \chatter relies on labels manually produced by malware analysts. This vetting process requires many man-hours and permits confident learning, although nicely integrated into the operation of \chatter. The man-hours, however, are not needed during the online operation. 

To capture the ordering of events in malware execution, \chatter uses \ngram{}s over abstract behavioral profiles. To boost its efficiency and achieve various desirable operational properties, \chatter focuses on one class of behavioral artifacts: network.  The use of $n$-grams in the context of malware classification is well established. However, {a novel contribution of our work is that it uses \ngram techniques to encode the order of subsequences of network communication events}. Our hypothesis is that each malware type has a unique communication pattern characterized by a certain order of events. This hypothesis is supported by Forrest et al.~\cite{forrest1996sense} which analyzed Unix system calls in a similar fashion. Building on this assumption we attempt to classify malware using only the order of network events. Our choice of network features is not arbitrary: as discussed in the rest of the paper, network features are cheaper than file system, registry, etc. features to obtain. For example, they can be captured without residing on the same host as the malware being executed. For our study we looked at three malware families: Zeus, Darkness, and Shady RAT, representing a diversity of malware intentions.

\BfPara{Contributions} 1) We introduce \chatter, a system for malware analysis and classification based on inexpensive order-based behavioral features. We argue for various deployment scenarios, and address operational needs.
2) We demonstrate the operation of  \chatter over various malware families using only the network interface for the source of behavioral artifacts. For this evaluation, we rely on manually produced labels. We demonstrate that \chatter is capable of accurately identifying malware family in a binary classification context. This is achievable  even when limiting the number of features below a quantity commonly used in the related literature.
3) We demonstrate an in-depth analysis of \chatter across multiple evaluation criteria and outline various open directions.

This paper builds on our work in~\cite{MohaisenWMA14}. While the paper at hand is almost entirely rewritten, and borrows very little text from~\cite{MohaisenWMA14}, it also incorporates new motivation, new threat model, use model, and preliminaries, new system description, and new and enhanced experiments and analysis. In particular, we updated sections~\ref{sec:introduction}, \ref{sec:related} and~\ref{sec:design}--\ref{sec:conclusion}, while adding~\ref{sec:prelim} to outline threat and use models. Technically, this work's extended contribution is towards methodical experiments and validation: unlike our prior work~\cite{MohaisenWMA14} where we study the performance of \chatter in isolation using fixed-length \ngram features, in this work we systematically evaluate \chatter with network features only, using  fixed length \ngram, combined grams, and utilizing feature selection for boasted performance (as explained in sections~\ref{metrics} and \ref{sec:results}; with performance gain of more than 20\% in accuracy). We finally evaluate our system over a dozen malware family; section~\ref{sec:data}, as opposed to three families in~\cite{MohaisenWMA14}.

\BfPara{Organization} The organization of this paper is as follows. We introduce the related work in section~\ref{sec:related}, preliminaries in section~\ref{sec:prelim}, the design of \chatter in section~\ref{sec:design}, its evaluation in section~\ref{sec:evaluation}, and a discussion in section~\ref{sec:discussion}. We sum up with concluding remarks in section~\ref{sec:conclusion}.

\section{Related Work}\label{sec:related}
The literature is rich with work on malware analysis and classification~\cite{tian2009automated,bailey2007automated,rieck2011automatic,park2010fast,tian2008function,rieck2008learning,kinable2011malware,ramilli2010multi,provos2007ghost,binsalleeh2010analysis,WestM14}. Broadly, the literature is divided into two schools of thought: signature based and behavior based techniques, with our work belonging to the latter, and most similar in nature to~\cite{rieck2011automatic,park2010fast,rieck2008learning,zhao2010malicious}. These works and others can be organized according to their relationship with techniques utilizing: machine learning for malware, general behavior-based analysis, memory signatures, network-related features, evasion prevention, $n$-grams, and event ordering. In the following, we elaborate on this research.

\BfPara{Machine learning for Malware} Machine learning techniques have been used to automate classification of codes and network traffic in the literature, with hundreds of related studies. The reader can refer to recent surveys in~\cite{SommerP10} and \cite{RossowDGKPPBS12} for the related work in this domain.

\BfPara{Behavior-based Analysis} 
There has been a large number of studies on using behavioral artifacts for malware analysis and classification~\cite{BerlinSS15,AndersonM16,bailey2007automated,MaricontiOACRS16,Firdausi10LEN}.
The work of Bailey et al. in~\cite{bailey2007automated} has motivated many of the related works on behavior-based malware classification. In~\cite{rieck2008learning,rieck2011automatic}, the authors use similar techniques for extracting features and leverage support vector machine (SVM) for classifying malware samples. Our work distinguishes itself in two respects. Although we share similarity with their high-level features, our system relies on the order of events, which exposes richer behavior. Second, we use analyst-vetted labels for evaluation, whereas the  other authors use heuristics over antivirus (AV)-returned labels.

%Our work is different from the prior literature both. First, we limit our attention to understanding and classifying individual families of malware sample, which is to the best of our knowledge a problem that is undermined in the literature. Second, to that end, our problem is limited in nature; we only use techniques that are designed for 2-classes classification problems, thus our error rates are smaller than those reported in the literature for multiple classes classification problems.

\begin{comment}
\BfPara{Reverse engineering} Kolbitsch et al.~\cite{KolbitschHKK10} introduced Inspector, which is used to automatically reverse engineer and highlight code sections responsible for ``interesting'' behavior. Related to that, Sharif et al.~\cite{SharifLGL09} proposed understanding code-level behavior by reverse-engineering code emulators. It is noteworthy that this and the previous work do not generate malware artifacts other than memory-related signatures, which have limited insight into characterizing generic samples.
\end{comment}

\BfPara{Traffic analysis} Related to our use of network features is a line of research on traffic analysis for malware and botnet detection. Such works include~\cite{JacobHKH11,GoreckiFKH11,botminer,GuZL08,gu2007bothunter,Wang17ZZYS}, with others paying particular attention to the use of fast-flux techniques~\cite{Spaulding18PKNM,HolzGRF08,NazarioH08,Skrzewski11,Spaulding18PKM,Spaulding17NM}. Support for our use of DNS features for malware analysis comes in~\cite{dns1,dns2,BilgeKKB11}. None of those studies IS concerned by behavior-based analysis beyond the use of remotely collected network features for inferring malicious activities and intention. Our system operates at network interface granularity to extract malware intelligence.

\BfPara{Evasion detection} Lanzi et al. introduced K-Tracer~\cite{LanziSL09} for extracting kernel malware behavior and mitigating the circumvention of loggers deployed in the kernel by rootkits. In~\cite{PerdisciLL08}, MacBoost is used for prioritizing malware samples by distinguishing benign and malicious code segments.  A system to prevent drive-by-malware based on behavior, named BLADE, is introduced in~\cite{blade}. A nicely written survey on such systems and tools is found in~\cite{Egele:2008}.

\BfPara{Leveraging $n$-grams} Using $n$-grams for malware classification is not new. However, work in the literature has looked at extracting features from executables (e.g., sequences of bytes in the binary files~\cite{rieck2011automatic}) or streams of communication traffic~\cite{wressnegger2013close}), but not a higher-level {\bf sequence of events} occuring while executing a malware sample. Other examples of low-level granularity attempts can be found in~\cite{PerdisciLL08,kolter2006learning,schultz2001data,wressnegger2013close}. Of particular interest is the concurrent work in~\cite{wressnegger2013close}, which derives $n$-gram network features for purposes of intrusion detection. Using network artifacts for identification of malicious activities, like botnets, is investigated in~\cite{StrayerLWL08,gu2007bothunter,botminer,GuZL08,Perdisci}. Further applications of characterizing malicious domain names using network traffic and artifacts (DNS queries, among others) are reported in~\cite{BilgeBRKK12,BilgeKKB11,provos2007ghost}.

Mariconti et al.~\cite{MaricontiOACRS16} proposed MaMaDroid, an Android malware detector that builds a behavioral model from sequences of abstracted application programming interface (API) calls that are performed by an app as a Markov chain. While MaMaDroid does not specifically use the \ngram{s}, it has the same effect in capturing the order of events (here API calls). Similar is the work of Shen et al.~\cite{shenandroid}, where $n$-grams are used over data flows to identify the sub-flows order.

{\noindent\bf Event ordering\hspace{3mm}} The basic idea of event order to characterize processes was first explored by Forrest et al. in their seminal work~\cite{forrest1996sense}. There, it was demonstrated effective for the detection of process-level intrusions. However, that work differs from ours in three respects: (1)~It is concerned with detection rather than classification, (2)~it uses system calls rather than networks features, and (3)~it uses whole sequences as a single feature that is easy to manipulate, rather than sub-sequences (as in $n$-grams) and their frequency.

%%%%%%%%%%%%%%%%%%%%%%%%%%%%%%%%%%%%%%%%%%%%%%%%%%%%%%%%%%%%%%%%%%%%%%%%%%%%%
%%%%%%%%%%%%%%%%%%%%%%%%%%%%%%%%%%%%%%%%%%%%%%%%%%%%%%%%%%%%%%%%%%%%%%%%%%%%%

\section{Preliminaries}\label{sec:prelim}

\subsection{Use Model}

\chatter adopts dynamic (network) analysis approach, as opposed to static analysis approaches, to classify malware using network traces that are collected by running the binaries in a sandboxed environment.   In other words, and in its operational mode, \chatter does not rely on host-based artifacts to characterize the behavior of malware, rather it passively collects network traffic that is generated from the hosts to synthesize malware behavioral patterns. Because this approach is less invasive, \chatter reduces the risks of intentional evasion whereby malware samples may attempt to inject unnecessary artifacts in their behavior to fool the dynamic analysis. Notice that the injection of such random artifacts, by either randomizing the behavior of the malware or inserting unnecessary behavior, is possible, and is addressed by Mekky et al.~\cite{MekkyMZ15}. 

\chatter consists of the bare-metal host(s) that run in a segregated (demilitarized) network that is connected to the Internet. A typical host would have processes associated with various activities including software updates, web traffic, and peer-to-peer traffic. Initially, these hosts are clean and only generate regular traffic, and each host is infected with only one malware for analysis and is reinitialized to clean state for every infection. In addition to the analysis hosts, a monitoring appliance is installed at key points on the network with a mirrored interface and network traffic is collected on the wire and analyzed to perform detection and classification of malware. To the observer on the wire, the whole traffic could be blended from multiple sources that include regular (background) traffic generated by the hosts and traffic generated by the malware samples. If the blended traffic is linearly mixed, then it is sent to a filter module that uses independent component analysis (ICA)\cite{MekkyMZ15} to separate malware traffic from background traffic. This step is taken before feature extraction and classification.

\subsection{Threat Model}
We assume an adversary in the form of a malicious software that is used to disrupt the operation of computer systems. \chatter is at heart a profiling system that relies on the behavioral artifact. Thus, while the adversary's goal is to remain stealthy, we assume a certain level of behavior that can be used for extracting features of such an adversary. For the purpose of this study, we assume that the behavioral artifacts generated by the adversary are unmixed with other sources of behavior. We assume that addressing such mixed behavior, where prevalent, is a secondary goal worth a separate investigation~\cite{MekkyMZ15}.

While the adversary could be aware of the defender's capabilities, including the machine learning techniques, the feature set, and the associated parameters, the adversary operates under certain constraints of achieving an end-goal, an attack that would ultimately result in behavioral artifacts used for its profiling. To this end, while the adversary may try to circumvent the system by manipulating the feature set to break the classifier, the defender would be aware of such a behavior to incorporate in further learning. Finally, of particular interest to our study is an adversary that relies on network artifacts for his operation. In particular, \chatter would operate best when such network artifacts are prominent in the behavior of the adversary. {Malware that does not use the network, e.g., air gap malware, falls out of the scope of this work. From our use and threat models, we exclude the case where malware uses tunneling of encrypted traffic. Despite continued increase of SSL/TLS-based malware families~\cite{ciscomalware,zscaler}, however, the majority of malware families (about 80\%) still do not rely on encryption, so we only deal with the samples which do not generate encrypted traffic in this work.

\subsection{The $n$-Gram Model}\label{sec:ngram}
In the following, we review the bases of the $n$-gram model. Interested readers should refer to~\cite{efros1999texture} for more details. 

Let $\mathcal{C}=\{c_1, c_2, c_3,\dots, c_\ell\}$ be a set of $\ell$ unique characters corresponding to the alphabet in a language $\mathcal{L}$. A collection of characters $c_1c_2c_3\dots c_s$, where $c_i \in \mathcal{C}$ for $1\leq i\leq s$, is said to be a sequence if this collection is ordered where repetition of elements is allowed. A word $w$ is a sequence $c_1 c_2 c_3\dots$, where the length of $w$ is the number of characters in it. A document $\mathcal{D}$ is defined as a collection of words $w_1w_2\dots w_d$. Based on that, $\mathcal{L}$ is defined as a system for communication based on words and a combination of them. Both characters and words are elements (tokens) in documents (strings). 

The $n$-grams are collections of adjacent elements in a string of tokens with the length of $n$. Notice that this definition is general, and it captures $n$-grams defined for both words and characters in robustly defined documents (strings) from a language $\mathcal{L}$. $n$-grams can be used to compute the probability of a token when the preceding token is given, using the chain of probability $P(c_1c_2c_3\dots c_s)$ defined as:
\begin{align}
\nonumber P(c_1)P(c_2|c_1)P(c_3|c_1c_2)\dots P(c_s|c_1\dots c_{s-1}) = \prod_{k=1}^s P(c_k|c_1\dots c_{k-1}).
\end{align}
For a bigram, we have $P(c_1^s) = \prod_{k=1}^s P(c_k|c_{k-1})$. Similarly, we define the probability approximation for $n$-gram as $
P(c_1^s) = \prod_{k=1}^s P(c_k|c^{k-1}_{k-n+1})$,  where the conditional probabilities are estimated using the relative frequency in $\mathcal{D}$ as:
\begin{align}
P(c_s|c_{s-1}) \approx \frac{C(c_{s-1}c_s)}{C(c_{s-1})}, P(c_s|c^{s-1}_{s-n+1}) \approx \frac{C(c_{c^{s-1}_{s-n+1}}c_s)}{C(c^{s-1}_{s-n+1})}
\end{align}
One way for building a feature vector for representing a certain abstraction using $\mathcal{L}$ is through the above probability estimation. Alternatively, we can calculate the probability directly by scanning through $\mathcal{D}$ to find all grams that meet the definition above for the given length $n$.  Defining words over behavioral artifacts is possible, as examined in~\cite{MekkyMZ15} using the same toolsets used in this paper. However, in this paper, and for the development of \chatter, we limit our attention to the case where behaviors are at the character level: we do not group those behavioral artifacts and consider the entire document $\mathcal{D}$ as a single word with as many characters as there is in the original behavioral profile.

\section{System and Design}\label{sec:design}
\chatter characterizes malware samples by executing them, using this intelligence for training purposes. For online operation, this capability is not required. We proceed by discussing \chatter{}'s design goals and then detailing how these are achieved in practice.

\subsection{Design Goals and Requirements}\label{sec:requirements}
In developing \chatter, we have several idealized functional and non-functional goals in mind. In particular, we develop \chatter so that it is cost-effective, less invasive, generalizable and multi-purpose, robust to behavioral changes, and accurate. In the following, we elaborate on each of those goals.

\BfPara{Cost effectiveness} Feature extraction should be computationally inexpensive, particularly in online operation. This cost effectiveness can be achieved by either limiting the number of classes of artifacts used for characterizing malware to a smaller set (e.g., network only, as opposed to fully-fledged systems such as AMAL~\cite{MohaisenA15}, which uses artifacts that pertain to network, memory, file system and registry usage) or by deriving shallow features across multiple classes. In \chatter, we emphasize the first approach by limiting ourselves to network features only. 

%We emphasize that \chatter ignores approaches requiring deep analysis of a large number of artifacts as in prior literature (e.g., AMAL~\cite{MohaisenA15} where the whole image of a 10GB hard drive and 256MB of RAM are analyzed to extract features).

\BfPara{Less-invasiveness} While possible to analyze malware in instrumented and virtualized environments, which is the approach we follow for building a baseline, it is desirable to collect such artifacts while running on the ``natural'' host OS. A system that can be deployed externally to observe malware is ideal. We keep in mind that this is only a design objective for the final system, not necessarily a requirement on how that system is arrived at.

\BfPara{Generalizable and multi-purpose} While the stated goal of \chatter is to characterize malware samples by their behavior, the system should be flexible enough for re-purposing outside the malware realm. In Section~\ref{sec:apps} we show two such applications that benefit from such generalized capabilities.

\BfPara{Robust to behavioral changes} Given that many malware families evolve over time to circumvent behavior-based techniques, one goal of our system is to resist this evolution by providing flexible and longitudinally-aware techniques.

\BfPara{Accurate} An ideal system for malware classification should aim to provide the greatest coverage while minimizing false positives. As such, when designing \chatter, we consider operationally acceptable accuracy within other optimizations (such as cost-effectiveness) of particular importance.

% With \chatter{}'s design goals in mind. we now show how they are achieved as we walk through the system design and subsequently, its evaluation.

%%%%%%%%%%%%%%%%%%%%%%%%%%%%%%%%%%%%%%%%%%%%%%%%%%%%%%%%%%%%%%%%%%%%%%%%%%%%%

\subsection{System Workflow and Operation}
\label{sec:flow}
\chatter{}'s overall workflow is visualized in Fig.~\ref{fig:flow1}. In describing this workflow we begin with our sandboxed execution environment and its output. We then describe how this output is transformed to make use of the $n$-gram technique to create a feature vector for each malware sample, which is then used for training a model to classify various species of malware in an automated manner. We note that any sandboxed execution environment (or bare-metal execution) can be used for extracting the features used in \chatter. We use our proprietary system named \automal, and described in greater depth in~\cite{amal}. \automal requires very little effort to be repurposed as the dynamic execution component of \chatter. \automal is a Windows-based system capable of collecting low-granularity artifacts speaking to how malware samples interact with memory, file system, registry, and network interfaces. \automal takes as input binaries that are likely to be malware. It consists of 4 components: submitter, controller, workers, and back-end storage. Input samples are queued until resources become available. The controller initiates virtual machines (VMs), loads configurations, and runs the malware sample. Once execution is completed the collected artifacts are logged in the back-end storage unit.

\subsubsection{Sandboxed Execution}\label{sec:sandboxed}
\chatter starts its operation by ingesting malware samples (provided by analysts as part of their research activities in security operations, customers interested in understanding classes of malware, or obtained from malware feeds of antivirus scanners~\cite{amal}), as shown in step 1 in Fig.~\ref{fig:flow1}. Given the limited resources and large number of samples at any point of time, our sandboxed execution environment (part of \automal~\cite{amal}) has a controller that checks for available VMs in the system (or a bare-metal units) for executing samples from the malware samples queue. If a VM (or bare-metal unit) is available, the controller runs this VM with a set of configurations provided by the analyst and executes the malware sample in the VM. The configurations are passed as a high-level script by the analyst: those configurations determine the operating system used in the VM, the virtual devices and resources associated with the VM, the software packages and programs installed in the VM, etc.). Upon executing the malware sample for a certain amount of time, as shown in step 2 in Fig.~\ref{fig:flow1}, our fully-fledged system collects a variety of dynamic execution features. Those include memory artifacts collected by various memory forensics and analysis routines (e.g., memory access, memory deletion, memory modification, size of reads and writes, utilized address space, etc.), file system artifacts (e.g., counts and content-related artifacts of files read, written, deleted and modified), registry artifacts (e.g., similar to the file system artifacts), and network artifacts. The amount of time a sample is executed is determined by an analyst based on several considerations, including a high-level notion of malware type and performance requirements

\begin{table}[t]
\begin{center}
\caption{Events used in composing the behavioral documents of malware samples.}\label{tab:features} 
\begin{tabular}{r|p{6cm}}
\toprule
Event class & Component events \\ 
\midrule
{\em IP and port} & unique dest IP, certain ports\\
{\em Connections} & TCP, UDP, RAW\\
{\em Request type} &  POST, GET, HEAD \\ 
{\em Response type} & response codes (200s through 500s)\\
{\em Size} & request (quartiles), reply (quartiles)\\
{\em DNS} & MX, NS, A records, PTR, SOA, CNAME \\
\bottomrule
\end{tabular}
\end{center}
\end{table}

Because \chatter only relies on one class of artifacts, namely network artifacts, AutoMal is repurposed where only network artifacts are considered, as shown in step 3 in Fig.~\ref{fig:flow1}. Those network artifacts are collected and indexed with the order of their appearance in the execution of the malware sample using timestamps. Among the network artifacts collected by \chatter, we mainly considered high-level header-based artifacts shown in Table~\ref{tab:features}. Those artifacts are mainly header-based, which means that they can be quickly obtained in real-time while executing the malware sample. We note that other deeper features from the contents (payload) of the network artifacts could also be utilized, although we avoid using such artifacts simply because due to efficiency and scalability reasons.

The artifacts in Table~\ref{tab:features} consist of multiple classes, including IP and port, connection (type), request and response (type), size information, and DNS-related features. For ports, we focus on a set of ports that are mostly utilized by malware samples in our analyzed feeds, some of which are generic services (e.g., ports 20, 21, 22, 25, 53, 80, 102, 110, 143, 389, 443, 465, 587, 636, 993, 995, 6665, 6347, 6679, 6697, and 8080 are used in our analysis).  The rest of the (type) artifacts are as highlighted in Table~\ref{tab:features}, whereas the size information is highlighted in quartiles (see~\cite{amal} for more details). In short, we define the size of a flow (request or response) as a number between 1 and 4 (inclusive). This number depends on which quartile the size of the flow fits within. The quartiles are computed over all the flows (requests and responses) associated with the execution of malware samples.

\begin{figure*}[!t]
\begin{center}
\includegraphics[width=0.85\textwidth]{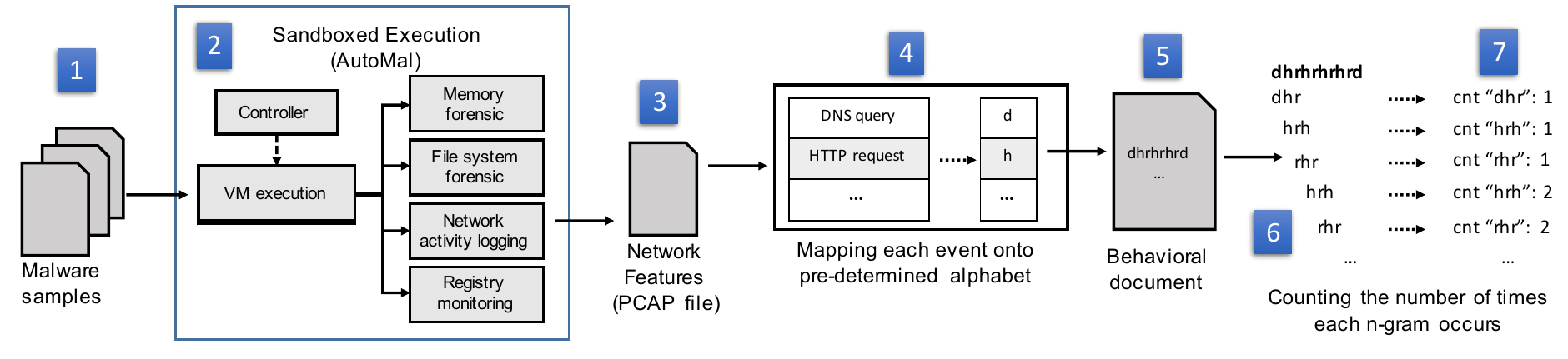}
\caption{\chatter{}'s sandboxed execution and feature extraction utilizing the network profiling in \automal and \ngram-based features extractor.}\label{fig:flow1}
\end{center}
\end{figure*}

\subsubsection{Behavioral Documents}
\label{sec:behavioral} \chatter abstracts the raw network behavior and artifacts obtained in step 3 in Fig.~\ref{fig:flow1} to compose a ``behavioral document.'' The behavioral document, as shown in step 4 in the same figure (for illustration purpose only), is simply an abstracted representation of the network artifacts to simplify their processing. To compose the behavioral document, we start with the network artifacts, and by fixing set of ``events'' (in advance) as shown in Table~\ref{tab:features}, we map those events into a set of unique characters in an arbitrary alphabet.  Then, using the network artifacts in step 3 we replace each event with the corresponding character to obtain a simplified representation of the network artifacts that preserve the order of the events. For example, in step 4 in Fig.~\ref{fig:flow1}, DNS query (of, say, MX type) is mapped to d, an HTTP request (of, say, type GET) is mapped to h, and so forth. By mapping the corresponding appearances of the events in the set in Table~\ref{tab:features} to the corresponding unique alphabet for the network behavior of a malware sample (and discarding unmapped behavior), we obtain the ``behavioral document'', as shown in Fig.~\ref{fig:flow1}. 

\subsubsection{\ngram Tokens} 
The next step in the operation of \chatter is the extraction of \ngram tokens from the behavioral document using the \ngram technique in section~\ref{sec:ngram}. By taking as an input a parameter $n$, \chatter calculates the tokens of length $n$ in the behavioral document, such that those tokens consist of adjacent characters in the behavioral document. For example, as shown in step 6 in Fig.~\ref{fig:flow1}, by starting with a behavioral document of {\tt dhrhrhrd}, and for $n=3$, \chatter calculates the following tokens as the output of the \ngram extraction component: {\tt dhr}, {\tt hrh}, {\tt rhr}, {\tt hrh}, {\tt rhr}, and {\tt hrd}. 

\subsubsection{\ngram Features} To extract features to represent a malware sample, we use the frequency of \ngram tokens obtained in the previous subsection and as illustrated in step 7 in Fig.~\ref{fig:flow1}. For instance, for the sample above, we build an index of the unique tokens and take the count over those tokens as the feature vector. Namely, for the same example, the index calculated over the tokens is [{\tt dhr}, {\tt hrh}, {\tt rhr}, {\tt hrd}, $\dots$] with a feature vector of [1, 2, 2, 1, $\dots$]. Note that the index is calculated over the superset of tokens that appear in the execution of all malware samples. Furthermore, an index and feature vectors could be constructed from the union of multiple grams (e.g., for an arbitrary $n$, we may consider the index of the feature vector as only the unique grams of length $n$, or all unique grams with a length less than or equal $n$; we elaborate more on the choice of the feature in the evaluation).

\subsection{Machine Learning Subsystem}\label{sec:ml}
The main purpose of \chatter is to provide means for automated malware classification using the behavioral features obtained through dynamic execution. To this end, \chatter supports various supervised machine learning algorithms for classification. Starting with a golden labeled dataset, \chatter builds a model utilizing the malware features extracted in the previous section. Upon validating the model using standard cross validation and accuracy measures, \chatter then uses the built model to classify malware in the wild to a family of interest. A simpler task that we also examine in the evaluation is the power of \chatter of detecting malware (this is, the classification of software into two classes of malicious and benign). In the following, we elaborate on the machine learning component of \chatter, including procedures and algorithms.

\subsubsection{Building Ground Truth} Devising methods for building ground truth as an open question in the security research community in particular and the machine learning in general. Methods that rely on manual inspection do not scale, whereas automated signature-based methods that utilize antivirus scanners and labels are susceptible to inaccuracies and incompleteness.  Fortunately, and as part of our security operations, manually-assigned names and labels of malware samples were given utilizing the variety of manual analysis, reversing, and memory signatures. While the method is costly, if it is to be used only for the operation of \chatter, such ground truth is obtained independently. We use this ground truth (more details in the evaluation). However, any other reliable ground truth source (that could potentially improve over antivirus scanners) can be utilized for the operation.

\subsubsection{Training and Validation} 
The task that \chatter tries to achieve is a binary classification: telling if a malware sample belongs to a malware family (or a class label) of interest or not. 
Given a set of malware samples represented by their feature vectors as extracted in the previous section and the corresponding labels for each sample, \chatter builds a model for the class of interest and other families (collectively represented as a second class; in malware detection, such a class corresponds to benign software). The model is trained (built) using part of the dataset and validated using the rest of the dataset. Independent of the algorithm used for building the model (detailed in section~\ref{sec:algorithms}), \chatter uses the $k$-fold cross-validation method for building the model, thus addressing the problem of overfitting.   For evaluating \chatter, we use the $k$-fold cross-validation method with $k=10$. In this method, the input dataset is divided into $k$-folds, where $k-1$ folds are used for training the machine learning algorithm and the remaining fold is used for testing. The process is repeated $k$-times by changing the testing dataset among the $k$ possible folds. At the end, the result of the classification algorithm  is computed as the average over the $k$ runs. We set $k=10$ due to its common use.

Upon training the model and testing it, and further establishing a confidence in the model, the underlying features, and the used parameters in the algorithm, we use the model in the wild for classifying (or detecting) malware. This is, we assign a label to a malware sample with the unknown label based on the determination made by our model for the given algorithm (details of how such determination is done is the testing phase of the algorithms in the next section; \ref{sec:algorithms}).

\subsubsection{Machine Learning Algorithms} \label{sec:algorithms}
Three machine learning algorithms are supported by \chatter: 1) the $k$-nearest neighbor (\knn), 2) support vector machine (SVM), and 3) decision tree classifiers. All three algorithms are intended for binary supervised learning and are capable of identifying the membership of a malware sample into one of two classes. Details on the operation of those algorithms are provided below for the completeness of our presentation of \chatter.

\BfPara{Support vector machines (SVM)} Given a training set of labeled pairs $({\bf x}_i, y_i)$ for $0<i\leq \ell$, ${\bf x}_i\in R^n$, and $y_i\in \{1,-1\}$, the (L2-regularized primal) SVM solves the following optimization: $\min_{{\bf w},b,\xi} \frac{1}{2}{\bf w}^T{\bf w} + C \sum_{i=1}^{\ell} \xi_i$ subject to $y_i ({\bf w}^T \phi({\bf x}_i) + b) \geq 1 - \xi_i$,  $\xi_i \geq 0$, 
%\begin{align}
%\nonumber \label{eq:svmprim}\min_{{\bf w},b,\xi}&\;\;\frac{1}{2}{\bf w}^T{\bf w} + C \sum_{i=1}^{\ell} \xi_i\\
%\nonumber \text{subject to }&\;\; y_i ({\bf w}^T \phi({\bf x}_i) + b) \geq 1 - \xi_i, \\
%\nonumber &\;\; \xi_i \geq 0
%\end{align}
where the training vectors ${\bf x}_i$ are mapped into a higher dimensional space using the function $\phi$, and the SVM finds a linear separating hyperplane with the maximal margin in this space. $C>0$ is the penalty parameter of the error term (set to 0.01 in this work). $\xi({\bf w}$, ${\bf x}, {y_i})$ is called the {\em loss} function, where we use the L2-loss  defined as $\xi({\bf w}, {\bf x}, {y_i})=\max(1-y_i{\bf w}^T{\bf x}_i, 0)^2$.

%{\noindent\bf\em Regularization.} The formalization in (\ref{eq:svmprim}) is called the L2-regularization. The L1-regularized L2-loss SVM solves the following primal problem (where $||\cdot||_1$ is the L1-norm):
%\begin{align}
%\min_{{\bf w}}&\;\;||{\bf w}||_1 + C \sum_{i=1}^{\ell} \max(1-y_i{\bf w}^T{\bf x}_i, 0)^2
%\end{align}

%{\noindent\bf\em Dual SVM.} The dual SVM problem~\cite{Hsieh:2008:dualsvm} is defined as: 
%\begin{align}
%\min_\alf f(\alf)=\frac{1}{2}\alf^T\bar{Q}\alf-\e^T\alf
%\end{align}
%subject to $0\leq \alpha_i\leq U$ for $1\leq i \leq \ell$, where $\bar{Q}=Q +D$, D is a diagonal matrix, and $Q_{ij}=y_iy_j\x_i^T\x_j$. For L1-Loss SVM, $U=C$ and $D_{ii}=0$ for $1\leq i \leq \ell$, while for L2-Loss SVM, $U=\infty$ and $D_{ii}=\frac{1}{2C}$ for $1\leq i \leq \ell$.
%%%should we talk about the dual form?
\BfPara{Decision trees} In evaluating \chatter, we utilize a single split tree for two-class classification using all of the features provided by \chatter. In particular, we use a small variation of the C4.5 algorithm. For a target class label $Y=y_1,\dots, y_n$ (here, $n=2$) and a set of feature vectors $\x_1\dots,\x_f$, at each internal node of the tree we apply a test to one of the inputs, namely $\x_i$, determining whether to go left or right in the tree branches based on the outcome of the test. When running over all of the training feature vectors, we mark the leaf nodes as the aggregate (mean) of all the training samples (to one of the class labels in $Y$). For testing, we do the same and assign the label of the leaf to that of the sample feature vector used to reach the leaf. There are variations of decision trees in the literature to provide better results (e.g., random forests). We did not try any of those techniques since this technique provided reasonable results. We leave integration of such techniques as future work.

\BfPara{The $k$-nearest-neighbor}
The \knn is a non-linear classification algorithm. In the training phase, we provide the algorithm two labels and a set of training samples (that are simply stored for the testing phase). In the testing phase, for each sample vector $\a$, we assign the label most frequent among the training samples nearest (using the Euclidean distance) to $\a$.   We refer the reader to a textbook explanation of the technique in~\cite{mlbook}.

\section{Evaluation}
\label{sec:evaluation}

To evaluate our work we begin by describing the malware samples utilized and how ground-truth is produced. This data is put through the \chatter workflow and feature vectors are computed. We then apply our machine learning technique of choice, producing evaluation metrics which are then interpreted with respect to system performance.

\subsection{Ground-truth and Labeling}
\label{subsec:data:groundtruth}
Labeling malware to establish a ground-truth is an important step in any supervised classification task. Prior literature relies heavily on family labels and names provided by anti-virus scanners. Several recent works~\cite{RossowDGKPPBS12,bailey2007automated,BayerCHKK09} have shown such scans to be unreliable. This is understandable to some extent, as these AV scanners are often designed with the primary goal of distinguishing malware from benign code -- not distinguishing between malware families.

%We note that, except in~\cite{mohaisen2013towards}, these studies do not particularly go in depth to analyze the limitations of the AV-based labeling but agree with us in total in this conclusion based on undisclosed experiments---yet, admitting it's a challenging problem with no other way to obtain labels, The works in~\cite{RossowDGKPPBS12,BayerCHKK09} use AV-labels as a ground truth. 

To this end, the malware samples we utilize have been collected over a considerable period of time, enabling expert analysts in Verisign's organization to manually identify and label them. This process is time-consuming; a previously unseen malware sample averages 10+ hours of manual characterization. However, family discovery is done in parallel with other important tasks. This time is not spent analyzing the binary just for this research initiative. Instead, Verisign analysts are contracted by customers to produce detailed reports on the modus operandi of a malware sample. 

In addition to customer submitted binaries, partnering anti-virus vendors and analyst's research endeavors also contribute to our malware repository. Yara signatures~\cite{yara} are sometimes applied to weed out irrelevant samples and a system called AMAL~\cite{amal} is applied to characterize samples from well-understood families. We emphasize that the majority of samples used in this study come from customers, where manual efforts are used for the labeling of samples. 
%\subsection{Dataset and Ground Truth}
%The dataset that we use for evaluating \chatter is shown in~\autoref{tab:dataset}, and is previously used in evaluating the work in~\cite{MohaisenA14b}. The dataset consists of 11 malware families, each of consisting of as few as 502 malware samples, to as many as more than 3500 malware samples. Each sample is given 

%%% provide the procedure used for giving malware samples their names.

%\begin{table}[t]
%\begin{center}
%\caption{Malware samples used in the evaluation of \chatter and their family association. Notice that the various families capture several types, which have varying levels of anticipated behavioral complexity.}\label{tab:dataset}
%\footnotesize
%\begin{tabular}{lll}
%\hline
%Malware family & \# & description   \\
%\hline
%Avzhan & 3458 & Commercial DDoS bot \\ 
%Darkness &  1878 & Commercial DDoS bot \\
%Ddoser & 502 &  Commercial DDoS bot \\
%Jkddos & 333 & Commercial DDoS Bot\\
%N0ise & 431& Commercial DDoS Bot\\
%ShadyRAT & 1287 & targeted; government and corps \\
%DNSCalc & 403& targeted; US defense companies\\
%Lurid & 399& initially targeted NGOs\\
%Getkys & 953& targets medical sector \\
%ZeroAccess & 568& Rootkit, monetized by click-fraud \\
%Zeus & 1975 & Banking, targets credentials\\
%\hline
%\end{tabular}
%\end{center}
%\end{table}

\subsection{Dataset and Malware Samples}\label{sec:data}
For the evaluation of \chatter, we use 11 malware families used in \cite{MohaisenA14b}: Avzhan, Darkness, Ddoser, Jkddos, N0ise, Shady RAT (SRAT), DNSCalc, Lurid, Getkeys, ZeroAccess, and Zeus. As we will soon describe, these families cover a wide range of network behavior. Our system is applicable for all malware families and our proof-of-concept selection is for demonstration purposes. Subsequent to the analysis described herein, we also ran our system on other families, including Ramnit, Bredolab, SillyFDC, and Virut. The systems' operation and accuracy during these larger trials were consistent for the results we present for our 11-family experiments.

Each malware sample in these families is obtained from an operational product, where sources of the malware samples include Verisign customers, partnering antivirus vendors, and researchers. Once the malware samples are fed into \automal they are executed for a fixed amount of time to generate artifacts. Table~\ref{tab:samplesize} reports on the quantity of the sample we virtualized and the average document length that resulted. To foster transparency and reproducibility of results, we release the dataset used in this work to the larger community. We note that the different families in our dataset belong to one of three classes: DDoS, targeted, or mass-market malware.

For every malware family set, we also have an equally-sized sample of random malware samples drawn from an expansive malware repository. 
%For example, our evaluation has 1025 Zeus samples. Thus when testing our model we also include 1025 non-Zeus samples consisting of many different malware families. 
In our proof-of-concept evaluation, we, therefore, treat family membership as a {\em binary classification task}. We now discuss each of these families in greater depth:

\begin{table}[t]
\begin{center}
\caption{Malware families used in the evaluation of \chatter, including set size and the average number of events per execution trace.}\label{tab:samplesize}
\scalebox{0.9}{
\begin{tabular}{r|rrl}
\toprule
    Family     & Quantity  &   Ch. Avg.& Description   \\
\midrule
Avzhan & 3458 & 70.31& Commercial DDoS bot \\ 
Darkness            &  1878    &  61.47&Commercial DDoS bot \\
Ddoser & 502 &  57.51& Commercial DDoS bot \\ 
N0ise & 431& 77.13& Commercial DDoS bot\\
Jkddos & 333 & 120.3& Commercial DDoS bot \\ 
Shady RAT            & 1287  &    52.74& targeted; government and corps \\
Getkys & 953&63.04&  targeted; targets medical sector \\
DNSCalc & 403& 82.37& targeted; US defense companies\\
Lurid & 399& 50.41&  targeted; initially targeted NGOs\\
Zeus             & 1975  &     50.74   &Banking, targets credentials\\
ZeroAccess & 568& 49.93 & Rootkit, monetized by click-fraud \\
\bottomrule
\end{tabular}}
\end{center}
\end{table}

\BfPara{$\diamond$ Zeus} One of the Trojans that attack financial sector through stealing information from the infected computers. To steal credentials, this Trojan hooks Windows API functions responsible for the communication between clients and the bank's website and modifies the returning results to hide its activities. 

%Zeus is a banking Trojan that targets financial sector by stealing credentials from infected victims. The malware steals credentials by hooking Windows API fucntions which intercepts communication between clients and bank's website and modifies the returning results to hide its activities.

\BfPara{$\diamond$ Avzhan} The DDoS botnet that was first reported in 2010~\cite{avzhan}. This botnet is similar to IMDDos, a Chinese process-based botnet announced by Damballa around September 2010~\cite{imddos}, where both families can be commercially hired to attack targets of interests. The owners of the botnet claim on their website that their botnet can strictly be hired to attack non-legitimate websites, for instance, gambling sites.

%is a ddos botnet, reported by Arbor Networks in their DDoS and security reports in September 2010~\cite{avzhan}. The family is closely related to the IMDDoS~\cite{imddos}, a Chinese process-based botnet announced by Damballa around September 2010. Similar to IMDDoS, Avzhan is used as a commercial botnet that can be hired (as a hit man) to launch DDoS attacks against targets of interest. The owners of the botnet claim on their website that the botnet can be used only against non-legitimate websites, such as gambling sites. 

\BfPara{$\diamond$ Darkness} a.k.a. Optima, is a commercial malware family developed by Russian criminals and released in 2009. As of the end of 2011, the 10th generation of this bot was released. This malware performs several functions, including launching DDoS attacks, stealing credentials and using infected machines for launching traffic tunneling attacks~\cite{optima}. 

%also known as Optima, is a malware family that is availably commercially and is developed by Russian criminals to launch DDoS, steal credentials and use infected hosts for launching traffic tunneling attacks (uses infected zombies as potential proxy servers). The original botnet was released in 2009, and as of end of 2011 it is in the 10th generation~\cite{optima}.

\BfPara{$\diamond$ DDoSer} a.k.a. Blackenergy, was reported and analyzed in 2007~\cite{ddoser}, is a DDoS malware that is capable of performing HTTP DDoS attacks. This malware is unique in the sense that it can target more than one IP address per DNS record. 

%Ddoser, also know as Blackenergy, is a DDoS malware that is capable of carrying out HTTP DDoS attacks. This malware can target more than 1 IP address per DNS record which makes it different than the other DDoS tools. It was reported on by Arbor networks and analyzed in 2007~\cite{ddoser}.

\BfPara{$\diamond$ JKDDoS} Malware that targets the mining industry~\cite{JKDDOS}. The malware was first reported by Arbor Networks in 2011, and its first generation was observed in September 2009.

%a DDoS malware family that is targeted towards the mining industry\cite{JKDDOS}. The first generation of the malware family was observed as early as September of 2009, and was reported first by Arbor DDoS and security reports in 2011. 

\BfPara{$\diamond$ N0ise} The DDoS tool used to recruit other bots to attack victims using HTTP, UDP, and ICMP flood, among other methods. Other functionalities of this tool include stealing credentials and downloading and executing other malware~\cite{n0ise}.

%n0ise is a DDoS tool with extra functionalities like stealing credentials from victim and downloading and executing other malware. The main use of n0ise is recruiting other bots to DDoS a victim using methods like HTTP, UDP, and ICMP flood~\cite{n0ise}.

\BfPara{$\diamond$ ShadyRat}  is a targeted malware that is used to steal sensitive information like trade secrets, patent technologies, and internal documents. The malware employs a stealthy technique when communicating with the C2 by using a combination of encrypted HTML comments in compromised pages or steganography in images uploaded to a website~\cite{shady}.

\BfPara{$\diamond$ DNSCalc} a.k.a. APT12,  is a malware is targeted towards research sector, where it steals sensitive information  through using the responses from the DNS request to calculate the IP address and the port number used for communication~\cite{dark}. 

%is a targeted malware that uses responses from the DNS request to calculate the IP address and port number it should communicate on, hence the name DNSCalc. The group is also known as APT12 by Mandiant. The malware steals sensitive information and targets research sector~\cite{dark}.

\BfPara{$\diamond$ Lurid} Malware that targeted US government and non-governmental organizations (NGOs), although it seems that there is no relationship between the targets indicators. This perhaps implies that the malware family is being used commercially as a hit man~\cite{lurd}. Three hundred attacks launched by this malware family were targeted towards 1,465 victims and were persistent via monitoring using 15 domain names and 10 active IP addresses.

%was first observed by the Japanese software security vendor Trend Micro on September 2011. Three hundred attacks launched by this malware family were targeted towards 1465 victims, and were persistent via monitoring using 15 domain names and 10 active IP addresses. While the attacks are targeted towards US government and non-government organization (NGOs), there seems to be no relationship between the targets indicator that perhaps the family is being used commercial as a hit man\cite{lurd}.

\BfPara{$\diamond$ Getkys} a.k.a. Skyipot, is a single-stage Trojan that targets aerospace, defense, and think tank organizations, through running and injecting itself in three targeted processes: outlook.exe, iexplorer.exe and firefox.exe. This Trojan communicates via HTTP requests and uses two unique and identifiable URL formats, such as the string ``getkys''~\cite{getkys}.

%(also known as Sykipot) is a single-stage Trojan that runs and injects itself into three targeted processes: outlook.exe, iexplorer.exe and firefox.exe. Getkys communicates via HTTP requests and uses two unique and identifiable URL formats like the string ``getkys." The malware targets aerospace, defense, and think tank organizations~\cite{getkys}. 
%\item {\bf t\_uglygorilla:}

\BfPara{$\diamond$ ZAccess} a.k.a. ZeroAccess, is a rootkit-based Trojan that targets most Windows OS and was reported first by Symantec in July 2011. The Trojan is mainly used to enable other malicious functions on the infected machine based on a pay-per-click advertising model. ZAccess is generally used to download other malware, open backdoors on the infected machines, among other functions~\cite{zeroaccess}.

\subsection{Feature Space and Space Reduction}\label{sec:features}
Running a malware sample in a sandboxed environment results in many artifacts and large PCAP files. Not all of these artifacts and the associated features are relevant nor meaningful in identifying a malware sample. Thus, we rely on expert {\em domain knowledge} to identify the network-related events of interest. To this end, Table~\ref{tab:features} highlights some of the 26 network-related events used in our analysis. Per Table~\ref{tab:featcount}, we observe that some events do not occur for some families (e.g., $n=1$) and far greater unique combinations of events never occur (e.g., where $n>1$ there is no permutation-scale growth). This makes empirical the advantages of using the sparse feature representation. For example, when $n=8$ our feature vector needs only $\approx$ 3,800 entries (the size of the feature space), rather than the $2.08^{11}$ entries needed for exhaustive representation of all combinations when $n=8$.
\begin{table}[t]
\begin{center}
\caption{The number of unique $n$-grams actually observed in each of the studied families for varying values of $n$.}\label{tab:featcount}
\scalebox{0.96}{\begin{tabular}{r|rrrrrrrr}
\toprule
Family~/~$n$&1&2&3&4&5&6&7&8 \\
\midrule
Avzhan&24&102&248&476&910&1506&2304&3252\\

Darkness&24&103&243&461&875&1503&2266&3149\\
Ddoser&26&107&250&491&900&1606&2636&3776\\
N0ise&24&106&249&480&891&1678&2264&3626\\

Jkddos&25&103&244&469&941&1703&2307&3122\\
SRAT&25&105&247&460&877&1536&2337&3300\\

Getkys&24&102&242&470&893&1659&2277&3364\\
DNSCalc&26&107&249&468&907&1578&2292&3379\\
Lurid&25&103&243&467&956&1539&2354&3130\\

Zeus&24&102&250&481&943&1690&2638&3794\\
ZeroAccess&26&108&250&489&913&1520&2306&3368\\
\bottomrule
\end{tabular}}
\end{center}
\end{table}
%Our baseline analysis to understand the impact and power of the $n$-gram based system 

%\begin{table}
%\begin{center}
%\caption{The number of unique $n$-grams actually observed in each of the studied families.}\label{tab:featcount}
%\footnotesize
%\begin{tabular}{r|rrrrrrrr}
%\toprule
%$n$ value&1&2&3&4&5&6&7&8 \\
%\hline
%Zeus&24&102&250&481&943&1690&2638&3794\\
%Darkness&24&103&243&461&875&1503&2266&3149\\
%SRAT&25&105&247&460&877&1536&2337&3300\\
%\bottomrule
%\end{tabular}
%\end{center}
%\end{table}

%24&105&249&479&880&1655&2413&3315\\
%25&103&243&497&902&1550&2395&3230\\
%25&104&249&464&900&1557&2268&3793\\
%%%%%%%%%%%%%%%%%%%%%%%%%%%%%%%%%%%%%%%%

\subsection{Evaluation Metrics and Scenarios}\label{metrics}

%To evaluate \chatter, we use evaluation metrics widely used in the literature~\cite{amal,BayerCHKK09}: the accuracy, precision, recall, and F1 score, which we define in the following. 

For a binary classification problem, in which it is required to determine if a given malware sample belongs to the class of interest $S$ or not, we define the following possibilities: 
(1)~True positives ($T_p$) are those samples correctly identified by the machine learning algorithm to belong to the class $S$. 
(2)~False positives ($F_p$) are those samples incorrectly marked by the machine learning algorithm to belong to $S$. 
(3)~True negative ($T_n$) are those samples marked by the machine learning algorithm correctly not to belong to $S$. 
(4)~False negative ($F_n$) are those samples incorrectly marked by the machine learning algorithm not to belong to $S$ (they are actually in $S$).  Using these outcomes and associated magnitudes, the precision, recall, accuracy, and F1 score are defined as:
\begin{align}
\text{Precision~(P)} &= {T_p}/{(T_p + F_p)},\\
\text{Recall ~(R)} &= {T_p}/{(T_p+F_n)},\\
\text{Accuracy~(A)} &= {(T_p+T_n)}/{(T_p+T_n+F_p+F_n)},\\
\text{F1 score~(F1)}  &= 2\times{(\text{P}\times\text{R})}/{(\text{P}+\text{R})}.
\end{align}
In the subsequent analysis, we scale all evaluation metrics to a percentage range by multiplying the result by 100.

\BfPara{Scenarios} In experimenting with \chatter, we consider the following scenarios. 1) Using all features resulting from the $n$-gram representation for a given $n$. 2) Using the combination of all features resulting from the $n$-gram, inclusive of all features with a gram of size less than or equal to $n$; this is, when a given $n$ is used the feature set would include, grams of size $n \dots 1$. 3) Using the top significant features selected using the recursive features selection (RFS).
RFS takes in two inputs, $k$ value, indicating the number of features desired, and $n$ features. RFS generates $n-1$ models by removing 1 feature from the $n$ features, then evaluates each $n-1$ model, then selects the best performing model. The process continues $m$ steps until $k=n-m$.  4) Using multiple algorithms on the same set of features to understand the impact of different algorithms on the performance of our classifiers.%  algorithm. %Our informal experiments with these techniques suggest little improvement in scalability or degradation of performance. This result is in line with a large body of literature on the problem~\cite{amal,BilgeKKB11,BayerCHKK09,maldimva,MohaisenA13}. 

%\begin{table*}[t!]
%\begin{center}
%\caption{Precision, recall, accuracy and F1-score for selected n-gram values}\label{tab:acrcy}
%\scalebox{0.88}{\begin{tabular}{|l||l||c|c|c|c||c|c|c|c||c|c|c|c||}
%\hline 
%&\sf{n-grams} & \multicolumn{4}{c}{1} & \multicolumn{4}{c}{4} & \multicolumn{4}{c||}{8} \\
%\hline
%& Algorithms& P & R & A & F1  
%& P & R & A & F1 
%& P & R & A & F1  \\
%\hline
%
%\multirow{3}{*}{\rotatebox{90}{Zeus}} &
%\knn  & 80.79 & 79.68 & 81.48 & 79.97 & 79.07 & 83.90 & 82.25 & 81.35 & 78.29 & 78.17 & 79.64 & 78.09 \\ 
%& SVM & 67.41 & 82.67 & 72.69 & 73.92 & 75.96 & 80.47 & 78.67 & 77.84 & 80.41 & 82.87 & 82.45 & 81.50 \\
%& Decision Trees  & 80.14 & 80.90 & 81.74 & 80.42 & 81.13 & 81.82 & 82.67 & 81.35 & 80.82 & 82.82 & 83.02 & 81.77 \\ \hline \hline
%
%\multirow{3}{*}{\rotatebox{90}{Dark.}} &
%\knn & 76.22 & 73.13 & 76.08 & 74.56 & 80.40 & 71.52 & 77.70 & 75.57 & 71.38 & 69.58 & 71.65 & 70.20 \\
%& SVM & 76.82 & 32.38 & 62.24 & 45.05 & 78.18 & 71.32 & 76.45 & 74.35 & 76.62 & 76.36 & 77.22 & 76.27 \\
%& Decision Trees & 80.45 & 72.56 & 78.20 & 76.07 & 81.75 & 72.89 & 79.04 & 76.93 & 80.50 & 68.37 & 76.39 & 73.59 \\ \hline \hline
%
%\multirow{3}{*}{\rotatebox{90}{SRAT}} &
%\knn  & 81.38 & 76.78 & 82.78 & 78.45 & 83.87 & 81.83 & 85.51 & 81.95 & 83.99 & 74.28 & 82.93 & 78.16 \\
%& SVM  & 76.88 & 65.43 & 75.88 & 69.55 & 83.70 & 82.94 & 86.23 & 83.03 & 85.68 & 80.86 & 86.33 & 82.71 \\
%& Decision Trees  & 85.16 & 81.11 & 86.44 & 82.60 & 88.28 & 81.65 & 88.01 & 84.45 & 86.13 & 78.92 & 85.54 & 81.85 \\ \hline \hline  
%\end{tabular}} 
%\end{center}
%
%\end{table*}

\begin{figure}[t]
\begin{center}
\subfigure{\label{fig:z3}\includegraphics[width=0.230 \textwidth]{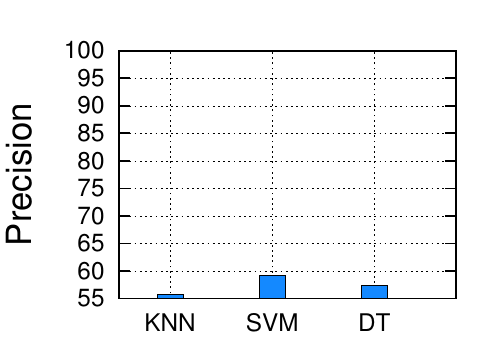}}
\subfigure{\label{fig:z4}\includegraphics[width=0.230 \textwidth]{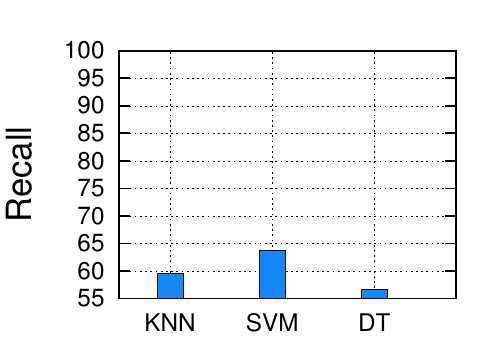}}
\subfigure{\label{fig:z1}\includegraphics[width=0.230 \textwidth]{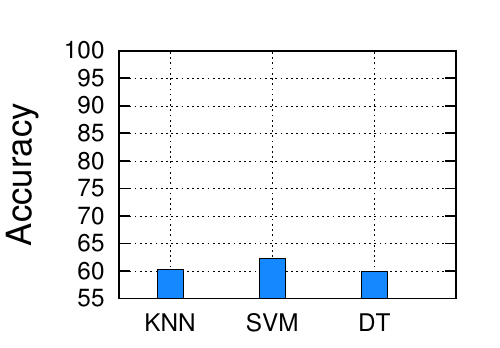}}
\subfigure{\label{fig:z2}\includegraphics[width=0.230 \textwidth]{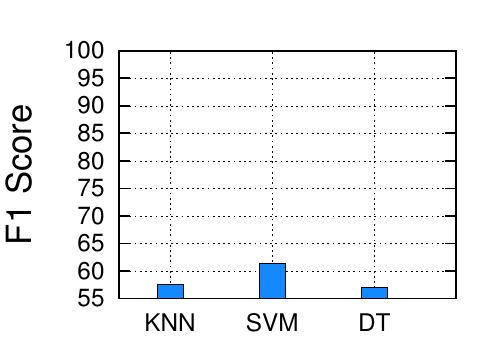}}
\caption{Performance measures for the Shadyrat malware family with network artifact classification using \chatter atop a baseline classifier}
\label{fig:dark}
\end{center}
\end{figure}

\begin{figure}[t]
\begin{center}
\subfigure{\label{fig:b1}\includegraphics[width=0.230 \textwidth]{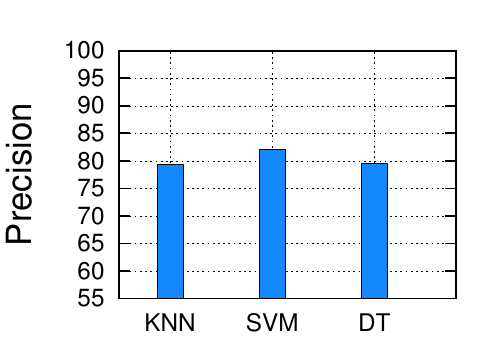}}
\subfigure{\label{fig:b2}\includegraphics[width=0.230 \textwidth]{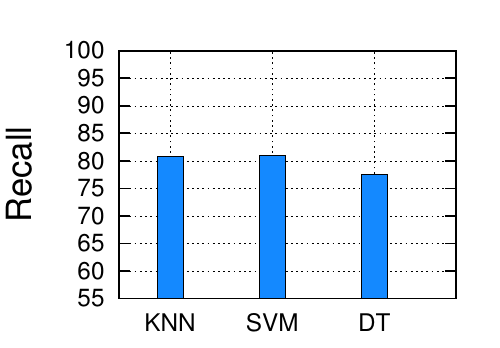}}
\subfigure{\label{fig:b3}\includegraphics[width=0.230 \textwidth]{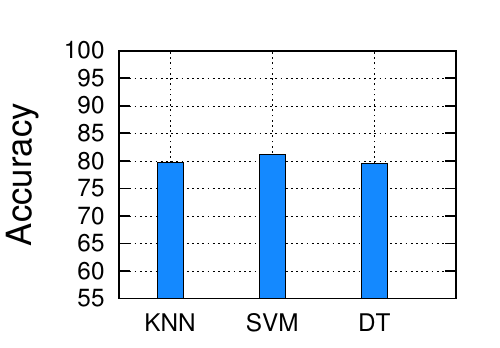}}
\subfigure{\label{fig:b4}\includegraphics[width=0.230 \textwidth]{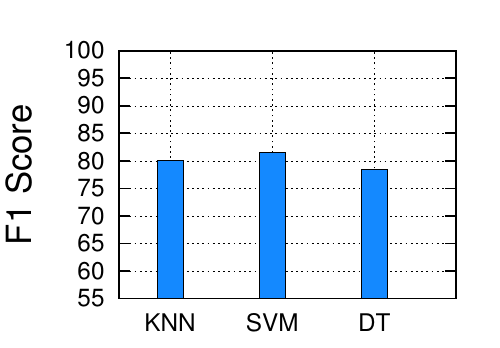}}
\caption{Performance measures for the Darkness malware family with network artifact classification using \chatter atop a baseline classifier}
\label{fig:bf}
\end{center}
\end{figure}

\begin{figure}[t] 
\begin{center}
\subfigure{\label{fig:d1}\includegraphics[width=0.230 \textwidth]{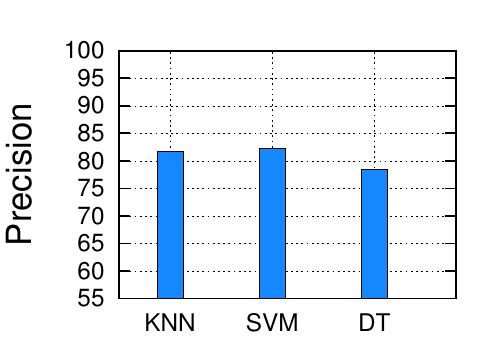}}
\subfigure{\label{fig:d2}\includegraphics[width=0.230 \textwidth]{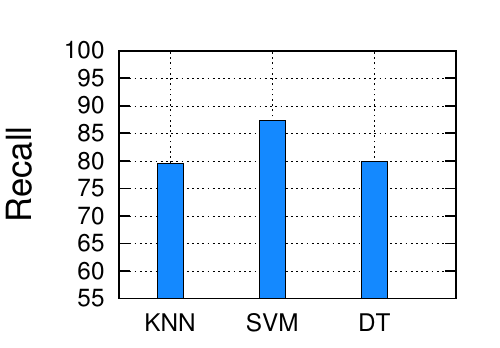}}
\subfigure{\label{fig:d3}\includegraphics[width=0.230 \textwidth]{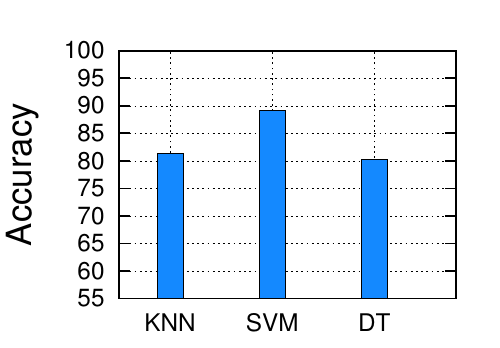}}
\subfigure{\label{fig:d4}\includegraphics[width=0.230 \textwidth]{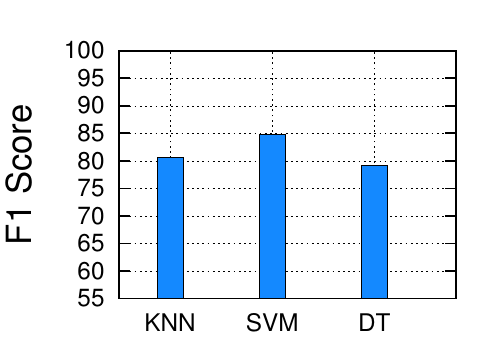}}
\caption{Performance measures for the Zeus DDoS malware family with network artifact classification using \chatter atop a baseline classifier}
\label{fig:zeus}
\end{center}
\end{figure}

%%%%%%%%%%%%%%%%%%%%%%%%%%%%%%%%%%%%%%%%

\subsection{Quantitative Results}
\label{sec:results}

We now present results from our evaluation. This section consists primarily of quantitative results, whereas later sections analyze and discuss these results in greater depth.

{\noindent\bf Results in isolation:} For this evaluation, we use three families that represent the three different malware classes in our dataset: Zeus (mass-market), Darkness (DDoS), and ShadyRat (targeted). The main purpose of this experiment is to understand the power of various machine learning algorithms and parameters when studying our datasets and preliminary features in isolation. Thus, we start with $n=1$, which corresponds to the bag-of-words method in the literature~\cite{MohaisenA15}, the three aforementioned machine learning algorithms, the four evaluation metrics, and the three malware families. The results are shown in Figs.~\ref{fig:dark}, \ref{fig:bf}, and~\ref{fig:zeus}. Among other observations, we notice the following. First, the performance, across all evaluation metrics, of the baseline feature is insufficient in almost all cases of malware families. Second, the performance of our baseline experiment varies greatly across families, and best results across all evaluation metrics is obtained with mass-market malware (Zeus), followed by DDoS malware (Darkness), then by targeted malware (ShadyRat), which is unsurprising pattern that holds for other malware families as verified through our analyses. Third, across all experiments, we notice that SVM provides the best results across all evaluation metrics, and the ranking of the two other algorithms is inconsistent and depends on the family studied. However, in all cases, the difference between the different algorithms is insignificant.   

Based on the baseline established in the first observation, we pursue the study of the impact of the $n$-gram features, and how they impact the performance of classification. Based on the third observation, we limit our attention to the SVM as the algorithm of choice when studying the impact of the $n$-gram parameters on the performance of classification. For the lack of space, we also only evaluate our system based on its accuracy. In the subsequent analysis, we perform classification for all of the eleven malware families in our dataset.

\BfPara{Evaluation using individual $n$-gram features} In this experiment, we focus on how using the new set of $n$-gram features affect the performance of the classifier. For this experiment, we use the SVM as the classifier of choice, for its better performance in the previous experiments. We study the performance over all families. In running  \chatter, we use individual features resulting from using the parameter $n$. Where $n$ is fixed, only features of  $n$ consecutive events are used in our feature vector. Results are shown in Table~\ref{tab:ngrams}. Based on those results, we make two observations. First, there is a monotonically increasing trend in the accuracy of the SVM classifier with the increase of the parameter $n$, even without features selection. This improved performance is clear across the board. Second, we notice, by experimenting with the 11 families in our dataset, a consistent ranking of the power of network-based features in general, and \chatter in particular, in profiling, characterizing, and detecting mass-market and DDoS families than the targeted families.  This insight is further highlighted in the ordering of three sample families belong to each class, in Fig.~\ref{fig:accu}.

\begin{figure}[!t]
\begin{center}
\includegraphics[width=0.45\textwidth]{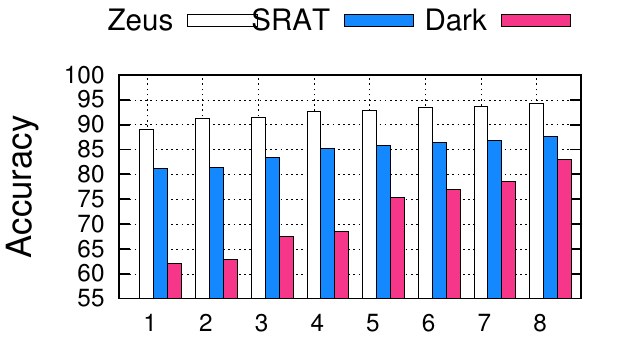}
\caption{The accuracy of classifying various malware families with varying  $n$.}\label{fig:accu}
\end{center}
\end{figure}

\begin{table}
\begin{center}
\caption{Accuracy of classifying various families for different values of $n$ when calculating the \ngram individual features.}\label{tab:ngrams}
\scalebox{0.93}{
\begin{tabular}{c|ccccccccc}
\toprule
family~/~$n$& 1 & 2 & 3 & 4 & 5 & 6 & 7 & 8 \\
\midrule
Avzhan &74.2& 74.8& 76.5& 76.7& 78.7& 83.5& 86.0& 91.2\\
Darkness &81.1& 81.4& 83.4& 85.2& 85.8& 86.5& 86.8& 87.6\\
Ddoser &84.1& 84.9& 85.6& 86.5& 92.8& 93.5& 93.7& 94.1\\
N0ise &88.0& 88.1& 88.8& 89.2& 90.7& 91.8& 91.8& 92.4\\
Jkddos &75.4& 76.1& 79.1& 85.2& 85.3& 85.6& 85.9& 87.7\\
ShadyRat &62.1& 62.8& 67.5& 68.6& 75.3& 79.0& 80.3& 82.9\\
Getkys &72.9& 72.9& 73.2& 74.8& 75.8& 77.0& 78.5& 79.8\\
DNSCalc &75.7& 77.0& 78.2& 78.4& 79.4& 81.3& 83.5& 84.4\\
Lurid &66.5& 67.3& 68.2& 69.7& 76.5& 77.5& 77.7& 85.7\\
Zeus &91.2& 91.2& 91.4& 92.6& 92.8& 93.5& 93.6& 94.2\\
ZeroAccess &83.3& 83.8& 84.6& 84.8& 89.7& 91.6& 92.9& 93.0\\
\bottomrule
\end{tabular}}
\end{center}
\end{table}

\BfPara{Evaluation using combined $n$-gram features} In this scenario, and for a given $n$, we compile a set of features for all combinations of events that are consecutive and are of the length less than or equal to $n$. For example, where $n=3$, we include features for lengths 1, 2, and 3. The results are shown in Table~\ref{tab:combined}. Overall, we notice an improved accuracy across the board in classifying the various families.% , and ranging from 1 to over 3\%

\begin{table}[t]
\begin{center}
\caption{Accuracy of classifying various families for different values of $n$ when calculating the \ngram combined features.}\label{tab:combined}
\scalebox{0.96}{
\begin{tabular}{c|cccccccc}
\toprule
family~/~$n$& 1-2 & 1-3 & 1-4 & 1-5 & 1-6 & 1-7 & 1-8 \\
\midrule
Zeus & 93.2& 93.5& 93.9& 94.0& 94.1& 95.3& 96.2\\
ZeroAccess & 84.9& 85.2& 85.3& 90.6& 94.1& 94.5& 95.2\\
ShadyRat & 63.2& 68.2& 71.2& 75.7& 79.4& 82.8& 85.8\\
Getkys & 74.7& 76.1& 76.7& 77.3& 78.2& 79.8& 82.6\\
DNSCalc& 78.3& 79.4& 79.9& 80.7& 84.2& 85.4& 86.0\\
Lurid & 69.0& 70.0& 71.2& 78.1& 78.9& 79.2& 88.4\\
AVZhan & 77.3& 79.1& 79.4& 81.3& 83.9& 87.5& 93.2\\
Darkness& 83.7& 83.9& 86.6& 86.8& 87.6& 88.7& 89.8\\
Ddoser & 86.4& 86.5& 88.6& 93.7& 94.5& 95.4& 96.0\\
N0ise & 89.3& 90.3& 90.5& 92.9& 92.9& 93.1& 93.6\\
Jkddos & 78.5& 81.0& 85.5& 85.5& 87.9& 88.5& 89.4\\
\bottomrule
\end{tabular}}
\end{center}
\end{table}

%Avzhan & 3458 & 70.31& Commercial DDoS bot \\ 
%Darkness            &  1878    &  61.47&Commercial DDoS bot \\
%Ddoser & 502 &  57.51& Commercial DDoS bot \\ 
%N0ise & 431& 77.13& Commercial DDoS bot\\
%Jkddos & 333 & 120.3& Commercial DDoS bot \\ 
%Shady RAT            & 1287  &    52.74& targeted; government and corps \\
%Getkys & 953&63.04&  targeted; targets medical sector \\
%DNSCalc & 403& 82.37& targeted; US defense companies\\
%Lurid & 399& 50.41&  targeted; initially targeted NGOs\\
%Zeus             & 1975  &     50.74   &Banking, targets credentials\\
%ZeroAccess & 568& 49.93 & Rootkit, monetized by click-fraud \\

\BfPara{Evaluation with selected features} In the following, we show the results with features selection over the previous experiment. In particular, we limit our attention to the 10\% top features in each family and compute the accuracy of the classifier for the different families. The results are shown in Table~\ref{tab:fselect}. Overall, we notice an improvement across the board. More importantly, we notice a final classification accuracy of 83\%-98.9\%, as compared to an initial accuracy of 62.1\%-91.2\% (with more than half of the families having the accuracy of less than 80\%). Notice that this improvement in accuracy is not the only improvement, but also includes a massive speed-up for the reduction in the number of features used for classification to only the most significant (10\%).

\begin{table}
\begin{center}
\caption{Accuracy of classifying malware families with varying $n$ values with recursive feature selection.}\label{tab:fselect}
\scalebox{0.93}{
\begin{tabular}{c|cccccccc}
\toprule
family~/~$n$& 1-2 & 1-3 & 1-4 & 1-5 & 1-6 & 1-7 & 1-8 \\
\midrule
Zeus& 94.2& 95.2& 95.8& 96.4& 96.5& 97.0& 98.9\\
ZeroAccess & 85.9& 85.9& 89.1& 93.4& 94.5& 94.5& 95.9\\
ShadyRat & 66.5& 69.5& 71.4& 78.9& 83.7& 84.5& 86.8\\
Getkys & 75.9& 76.2& 76.4& 77.7& 80.8& 81.6& 83.0\\
DNSCalc & 78.9& 80.2& 81.8& 82.4& 82.6& 85.6& 86.2\\
Lurid & 70.0& 71.7& 74.4& 77.2& 80.0& 82.1& 89.6\\
AVZhan & 78.3& 79.2& 81.9& 81.9& 87.4& 88.4& 94.9\\
Darkness & 86.8& 87.8& 87.9& 88.6& 89.4& 90.4& 90.9\\
Ddoser & 88.3& 89.9& 91.8& 96.2& 96.8& 97.0& 97.3\\
N0ise & 91.3& 91.9& 92.8& 93.8& 94.4& 96.6& 97.2\\
Jkddos & 79.1& 81.1& 88.1& 88.4& 89.4& 90.3& 93.6\\
\bottomrule
\end{tabular}}
\end{center}
\end{table}
\section{Discussion}
\label{sec:discussion}

When $n=1$ ordering plays no role, whereas there are tremendous ordered dependencies when $n=8$ (our upper limit). We observe that malware families tend to peak in performance when $n=8$. This observation is further validated when combining features and feature selection. The accuracy graphs in Figs. \ref{fig:dark}, \ref{fig:bf} and ~\ref{fig:zeus} show an upward trend as $n$ grows. We conclude that order does improve classification by about 7\% to 23\%.

%We note that the classification for the Darkness/DDoS malware family was less accurate than with other families, even when considering baseline features. This speaks to the way Darkness infects a host machine and uses it as a bot. Darkness creates no files (which would trigger many of our baseline features) and does most of its characteristic manipulations inside the registry. While possible to do order-based evaluation of registry events, having the access to monitor a host registry in real-time presents privacy and complexity hurdles which are not present in a network-only system.

%%%%%%%%%%%%%

\subsection{Meeting Design Requirements}\label{sec:meetingreq}
In the following we highlight how \chatter meets the design requirements outlined in Section~\ref{sec:requirements}:

\BfPara{$\diamond$ Cost-effectiveness} \chatter{}'s cost-effectiveness is two-fold: (1)~It uses only a single class of (network) artifacts, and (2)~It abstracts features from curated event traces rather than raw interface dumps. Prior literature has shown a system characterizing malware using only network artifacts can run an order of magnitude faster than a system that looks at a large spectrum of features. For example, AMAL~\cite{amal} is a fully-featured infrastructure that ideally utilizes 128 virtual machines towards processing 23,000 malware samples daily. \chatter, on the other hand, could process 370,000 malware samples per day using the same infrastructure. This number significantly exceeds the sample quantity Verisign's operations receive and analyze on a  daily basis. Indeed, the number is also greater than the 250,000 samples a popular antivirus provider like Sophos detects and analyzes daily~\cite{ciscox}.

\BfPara{$\diamond$ Less-invasiveness} Using network features makes \chatter less invasive. The network events can be gleaned on the network without having to reside on the host machine. However, this capability is also susceptible to noise from other system processes using the network interface in parallel with the malware sample. This could be limited by running the malware and the host in a monitored mode of operation. 

\BfPara{$\diamond$ Generalization and flexibility} \chatter{}'s ability to evolve with malware binaries is straightforward given its operational context. Because novel malware is being created, analyst efforts continue to be brought to bear on samples' reverse-engineering and analysis. Family labels are a cheap side-effect of this ongoing demand, resulting in a wealth of expert-annotated and longitudinal ground-truth that can be utilized via periodic retraining.

\BfPara{$\diamond$ Accuracy} In isolation, \chatter{}'s accuracy is less than that of full-featured systems, e.g., AMAL~\cite{amal}. However, we still contend that our performance is \textit{operationally acceptable}. The contexts in which one needs to make a family classification are very different when determining the presence of malware. Remember also that accuracy can be improved by using more expensive techniques (possibly as a second-pass if \chatter{}'s models indicate low classification confidence). This work is also interested in the trade-off between complexity, accuracy, and operational costs.

%discussion points:
% 1. Why did SVM perform bad with n=1 and outperformed the other algorithms as n increased?
% 2. Explain why FS features combined with n-gram features improve the score
%  2a. Darkenss didn't perform well because it only creates one file and several registry entries to install itself as a service. The registry artifacts are not consideered in the classification study
% 3. Trade off between high accuracy and cheap artifacts. Overall the order does improve the classification by more than 5% as n gets larger for the n-gram method, but the optimal point is n=4 or 5.
% 4. 

\subsection{System Limitations}
\label{sec:limitations}

We acknowledge various shortcomings of \chatter, and examine how a knowledgeable attacker could utilize gamesmanship to circumvent our classification strategy. 

\BfPara{Noised features} As with most behavior-based systems for malware classification, \chatter performs best when malware samples do not produce extra information to disguise their behavior and manipulate machine learning algorithms. However, unlike systems that make use of exact matching of behavior profiles, \chatter provides some flexibility and robustness in the grouping patterns based on $n$-gram features.  The problem is a generic one which we address in two ways.  First, we emphasize that not all the features generated by a malware sample need to be used by learning algorithm: a feature selection algorithm can be used to reduce the impact of the noised features. Second, regardless of injected noise, certain events in the operation of a malware sample must  happen in the same partial order. Our future work to address this limitation is to derive features of those events as they happen in their partial order by filtering out the noise between them. While this might seem to require a deep understanding of the studied malware families and their expected behavior, well-understood signal processing techniques show potential in this domain. %For example, in the future it is worth considering how statistical and information theoretical characteristics of the \ngram features can guide the process of deriving representative and meaningful features.

\BfPara{Adaptive malware} Certain forms of malware are capable of changing their behavior based on the environment in which they are run. This creates issues for \chatter as it does for other behavior-based sandboxes. We address this in two ways:

First, \automal, the sandbox environment for \chatter generates patches to deceive malware samples by providing registry values indicating execution is occurring on bare metal. Second, for sophisticated malware that does not respond to those patches, malware samples are actually run on bare metal (or using hardware virtualization). Of course, this a problem specific to the generation of behavior profiles, whereas actual operation is unaffected by such manipulations.

{\noindent\bf\em Continuous training and cost of labeling:} Because of the evolution of malware samples, continuous training is needed in our system to adapt to changes in artifacts generated. While this issue might seem an inherent shortcoming for machine learning based techniques, it is addressed naturally in \chatter. As mentioned earlier, many of the malware samples fed into \chatter belong to customers and require reverse engineering, deep analysis, and manual inspection. To that end, this process provides a natural venue for obtaining features, labels, and training sets for \chatter. %Notice that the frequency of retraining in our system is less often than  

Our objective in this work is characterizing a threat model against \chatter by defining ways where an adversary can alter the pattern of network traces when running malware samples in a dynamic analysis environment. An adversary may alter the sequence of traces in various ways hoping to confuse monitoring systems and to make the extracted features less discriminatory, either by injecting new packets into the network or modifying existing ones or dropping packets in transit if they can intercept them.

We argue that sensitivity of classification in \chatter is affected by environmental noise or by malicious actions intended to poison traces of execution using operations like: addition, deletion, and modification of behavioral events. Consequently, an adversary who executes these operations seeks to record distorted patterns as input to the classifier and therefore hopes to evade detection by generating faulty results. We identify potential attacks that can target the classifier and affect  the overall performance of \chatter. Each of those attacks, while ranging in sophistication, rely on unique and basic features of the proposed approach for discriminatory feature extraction. These attacks can be modeled by simple linear combination attacks or they can be more sophisticated such as non-linear or spatial attacks.

\BfPara{Linear attacks}  With linear attacks, the source of the attack can be modeled as a linear combination of background traffic generated from regular hosts (based on web browsing, software updates, \dots etc) with network events generated by the malware sample. Noise in the behavioral sequence can affect the classifier's performance and to overcome this requires either applying a proper filter or use of tight monitoring in sandboxed environments. However, malware samples may actually simulate linear noise by injecting unnecessary artifacts in their behavior to fool the dynamic analysis, thus mimicking background traffic and evading detection by simulating multi-blended traffic. Methods such as independent component analysis (ICA) can be used to separate linearly combined traffic by analyzing distributions of features extracted from mixed traces into two estimated distributions, provided that traffic sources are statistically independent. 

\BfPara{Non-linear attacks} Unlike linear attacks which assume traffic is linearly mixed,   spatial attacks are based on non-linear combinations of the features that exhibit spatial distortion properties.  Two techniques seemed promising for defending against spatial attacks, namely: sequence alignment and skip-grams. Sequence alignment is a well-established method in bioinformatics used for identifying regions of similarity between two sequences that might be a consequence of functional, structural or evolutionary relationships. Typically, a ground truth of clean sequence is used to align various noisy sequences. On the other hand, skip-grams can be used to model noisy sequences by finding the pattern of characters (or words) that do not have to be consecutive.  For example, a 1-skip grams algorithm would be applied the same way as 2-grams, but by skipping a word in every two grams in the analyzed document. They are considered a generalization of $n$-grams model which can be used to overcome noise in the sequence.

\subsection{Other Applications}\label{sec:apps}
While the main application we used in \chatter relies on transforming behavioral profiles into documents and using them for understanding the behavior of malware utilizing \ngram techniques, the concept is generic and can be applied to a wide variety of applications. In the following we identify several potential applications which can benefit from \chatter: (1)~{\em Process-based DDoS detection:} While our system studies a specific DDoS malware family, our system can be generalized to understand any process-based DDoS attack by observing traffic on the wire, generating sufficient artifacts that can be used to derive features and footprint such attacks. (2)~{\em Advanced persistent threats:} Often such threats (process-based) result in many artifacts that are generated over a long period of time, rendering research systems less effective in characterizing them. One potential area of improvement is to rely on the inter-event patterns they generate, using \chatter.

One potential application to the technique is what we term as the ``adversary incubation'', in which an attacker is led into a controlled environment and observed over a long period of time. The behavior of the adversary can be then used to characterize him and using his network activities in particular. We notice that the adversary incubation is one step for understanding the ``advanced persistent threat'' (APT), and is widely used in the industry. The incubation usually result in network artifacts generated by the adversary and spread over a long period of time, as a result of low rate activities that the adversary hope to go unseen by anomaly detection defenses. Utilizing systems like AMAL for understanding the adversary might not be as effective, whereas using the inter-event patterns of actions taken by the adversary may reveal valuable about the attacker and help attributing attacks to him.

\section{Conclusion}\label{sec:conclusion}
Motivated by the need for deriving new and easy-to-obtain features, we introduced \chatter, a behavior-based malware classification system. \chatter uses behavioral artifacts generated by malware samples at runtime to characterize malware. At its core, \chatter considers the order in which behavioral events occur. We notice that order-based features can be captured and analyzed using the \ngram technique widely used in document classification. With its many advantages enumerated in Section~\ref{sec:design}, and using three malware families, \chatter is shown to be reasonably accurate at classifying malware samples into their respective families. 

This paper considered order-based behavioral features for classification of malware samples in its simplest form. Addressing the limitations outlined in Section~\ref{sec:limitations} is the immediate future work. In particular, we would like to explore partial-order features for fingerprinting malware samples. Those features would address noised features (both intentional obfuscation by malware authors, and unintentional, due to mixed signals on-the-wire). Realizing the applications listed in Section~\ref{sec:apps} using the \chatter methodology is another future work that we would like to explore.

\BfPara{Acknowledgements}
An earlier version appeared in IEEE CNS~\cite{MohaisenWMA14}. Part of this work was done while A. Mohaisen and O. Alrawi were at Verisign Labs. This work was supported in part by the Global Research Lab. (GRL) Program of the National Research Foundation (NRF) funded by Ministry of Science, ICT (Information and Communication Technologies) and Future Planning (NRF-2016K1A1A2912757).

\balance
%\bibliographystyle{icstnum} 
%\bibliography{sigproc}

\end{document}